\pdfoutput=1
\RequirePackage{ifpdf}
\ifpdf % We are running pdfTeX in pdf mode
\documentclass[pdftex]{sigma}
\else
\documentclass{sigma}
\fi

\numberwithin{equation}{section}

\begin{document}

\allowdisplaybreaks

\renewcommand{\PaperNumber}{003}

\FirstPageHeading

\ShortArticleName{Supersymmetric Proof of the Hirzebruch--Riemann--Roch Theorem}

\ArticleName{Supersymmetric Proof of the Hirzebruch--\\ Riemann--Roch Theorem for Non-K\"ahler Manifolds}

\Author{Andrei V.~SMILGA}

\AuthorNameForHeading{A.V. Smilga}

\Address{SUBATECH, Universit\'e de Nantes, 4 rue Alfred Kastler, BP 20722, Nantes 44307, France\footnote{On leave of absence from ITEP, Moscow, Russia.}}
\Email{\href{mailto:smilga@subatech.in2p3.fr}{smilga@subatech.in2p3.fr}}

\ArticleDates{Received November 10, 2011, in f\/inal form January 04, 2012; Published online January 08, 2012}

\Abstract{We present the proof of the HRR theorem
for a generic complex compact mani\-fold by evaluating the functional integral
 for the Witten index of the appropriate super\-symmetric quantum mechanical system.}


\Classification{53C55; 53C80}

\section{Introduction}

It is fairly clear now that the theoretical high-energy physics (at least, its part represented in the
hep-th section of the Archive), being unfortunately deprived since a certain time of the
experimental feedback, is gradually transforming into a branch of pure mathematics. The synthesis of these
two sciences is fruitful for both of them: f\/ield theorists have absorbed and use a lot of mathematical
techniques and methods, but also many physical concepts turned out to be very useful in treating pure
mathematical problems. One can e.g.\ mention in this regard the paper \cite{Witknot} where the methods
of quantum f\/ield theory were used to derive topological knot invariants.

Another distinguished example is the Atiyah--Singer theorem for the index of Dirac operator which was f\/irst
proven by pure mathematical methods \cite{AS} and then a physical, in many respects more simple
and clear proof of this theorem has been found \cite{A-GFW}. This physical proof was based on the isomorphism
of the classical complexes (de Rham complex, Dolbeault complex, Hirzebruch complex and Dirac complex)
to certain supersymmetric quantum mechanical systems, with the indices of all these complexes being mapped
to the Witten indices \cite{Witind} of the proper SQM systems.

The Atiyah--Singer theorem was not, however, proven in these papers in its whole generality. Namely,
the formula for the index of the Dolbeault operator was derived for K\"ahler manifolds but not for a generic complex
manifold. In the present paper, we f\/ill this gap.

Before going into details, let us brief\/ly outline the structure of the proof.
 \begin{itemize}\itemsep=0pt
\item At the f\/irst step, we write down, following \cite{my}, the SQM system whose supercharges may be interpreted as
exterior holomorphic derivative~$\partial$ and its conjugate $\partial^\dagger$ (see equations~(\ref{start}) and~(\ref{W}) below).
\item For K\"ahler manifolds, the functional integral for the index $I={\rm Tr}\big\{(-1)^F e^{-\beta H} \big\} $ can be
reduced in the semiclassical limit $\beta \to 0$ to the ordinary integral (\ref{indBis}). In the K\"ahler case,
this  coincides with the integral representation (\ref{indpoAS}) used usually by mathematicians.
 \item For the manifolds that are not K\"ahler (such that the K\"ahler form $\omega$ is not closed), but which
satisfy the so called SKT condition, $\partial \bar\partial \omega = 0$, the functional integral can as well
be reduced to the ordinary integral (\ref{indBis}). (The integral representation (\ref{indBis}) for the Dolbeault
 index of the SKT manifolds was derived earlier in \cite{Bismut} by purely mathematical methods.) In this case,
 the integrands in  (\ref{indBis}) and (\ref{indpoAS}) {\it do} not coincide. One can show, however, that their
dif\/ference is an exact form such that the integrals (\ref{indBis}) and (\ref{indpoAS}) are equal.
 \item For generic complex manifolds, a direct evaluation of the functional integral is dif\/f\/icult by the reasons
explained below. One can notice, however, that the Dolbeault complex is equivalent to a Dirac complex  involving
extra torsions (in the K\"ahler case, the torsions are absent). We show then that one can actually {\it unwind}
these torsions by deforming continuously the lagrangian in such a way that supersymmetry is kept at every value
of the deformation parameter. The index is not changed under such a deformation and coincides thereby with the
index of a certain torsionless Dirac operator. The functional integral for the latter can be reduced to an
ordinary integral (\ref{indSmil}) by standard methods. One can proceed then as in the SKT case and observe that
the integrands in equations~(\ref{indSmil}) and~(\ref{indpoAS}) dif\/fer by an exact form.
  \end{itemize}

\section{Dolbeault complex}

We remind here some basic facts and def\/inition of the complex dif\/ferential geometry.

We precede it with the following remark. As was mentioned above, theoretical high energy physics  is
merging now with mathematics, but at the moment this merge is far from being complete. As a result, there exist
now two dif\/ferent communities with mathematical and physical backgrounds, studying in many cases very similar or
just identical objects. Two communities use two dif\/ferent languages, such that a mathematical paper
is more often than not incomprehensible to a physicist, and vice versa.

Speaking of this particular paper, it is written by a physicist and is addressed mainly to physicists
even though its subject is in fact pure mathematical. Thus, I will use whenever
possible the physical terminology even in the cases when a translation to mathematical language exists (and is known to
the author). For example, we will not talk about {\it line bundles} and by no means about {\it sheaves of germs}, but
only about {\it Abelian gauge fields}. In some cases, mathematical translations will be given in footnotes.

A {\it complex manifold} is a $2n$-dimensional manifold covered by overlapping regions $U^{(a)}$, with every region being
described by a set of complex coordinates $z^j$, $j = 1,\ldots,n$ with the metric having Hermitian form
\begin{gather}
\label{metric}
g_{JK}   =  \begin{pmatrix} 0 & h_{j\bar k} \\ h_{k \bar j} & 0 \end{pmatrix}.
 \end{gather}
The reality of the metric implies $(h_{j \bar k})^\star  = h_{k \bar j}$ such that
$ds^2   =   2h_{j \bar k} dz^j d\bar z^{\bar k}$. The coordinates~$z^{j(a)}$ and~$z^{j(b)}$ in the
overlap $U^{(a)} \cap U^{(b)}$ are expressed into one another by holomorphic functions,
$ z^{j(a)}   = f^{j(ab)}(z^{k(b)})$. The {\it Dolbeault complex}
 is a set  of all purely holomorphic forms,
\begin{gather}
\label{Ap}
A^{(p)}   =  A_{j_1 \ldots j_p} dz^{j_1} \wedge \cdots \wedge dz^{j_p}.
 \end{gather}
These forms should be regular on the manifold meaning that their norm is bounded
 \begin{gather}
\label{norm}
 \big\|A^{(p)}\big\|^2    =  A_{j_1 \ldots j_p}  A^\star_{\bar k_1 \ldots \bar k_p} h^{\bar k_1 j_1} \cdots h^{\bar k_p j_p}   < \infty   .
 \end{gather}

Consider the operator of  holomorphic exterior derivative, $\partial A^{(p)} = \partial_{k} A^{(p)}
dz^k \wedge dz^{j_1} \wedge \cdots \wedge dz^{j_p}$ and its conjugate $\partial^\dagger$.
The operators
$\partial$ and $\partial^\dagger$ are nilpotent and can be interpreted as the supercharges~$Q$,~$\bar Q$. Their anticommutator
$\{\partial, \partial^\dagger \}$ is called {\it Dolbeault Laplacian}  and can be associated with the Hamiltonian.

A {\it  K\"ahler manifold} is the manifold where the metric $h_{j \bar k}$ satisf\/ies the condition
$\partial_l h_{j\bar k} =  \partial_j h_{l\bar k}$ and can hence be derived from a {\it K\"ahler potential},
$h_{j\bar k} = \partial_j \partial_{\bar k}   K$.
Alternatively, one can say that the K\"ahler form $\omega = h_{j \bar k}   dz^j \wedge d\bar z^{\bar k} $ is closed, $\partial \omega =
\bar \partial \omega = 0$. For K\"ahler manifolds, Dolbeault Laplacian coincides with the conventional Laplace--Beltrami operator.
In a generic case, it does not.

Note also that, in the K\"ahler case, the  ``small'' Dolbeault complex just described can be enlarged. The
 ``large'' Dolbeault complex involves all forms
(not necessarily holomorphic) and, on top of $\partial$ and $\partial^\dagger$, also  antiholomorphic exterior
derivative $\bar \partial$ and its conjugate $\bar \partial^\dagger$. The SQM systems isomorphic to large  Dolbeault complexes
enjoy extended supersymmetry. They will not interest us in this paper.

Complex manifolds can be described using real notations if introducing the so called {\it complex structure} matrix $I_M^{\ N}$,
$M \equiv \{m, \bar m \}$. It represents a tensor satisfying the conditions
 \begin{gather}
\label{I}
I_M^{\ N}   I_N^{\ P}   =  -\delta_M^P , \qquad  I_{M}^{\ P}   g_{PN} + I_{N}^{\ P}   g_{PM}   \equiv  I_{MN} + I_{NM}   =  0  ,
 \end{gather}
where $g_{MN}$ is the metric.
Once complex coordinates $z^j$, $\bar z^{\bar j}$ and the Hermitian metric $h_{j \bar k}$ are def\/ined, the complex structure
matrix has only  nonzero components with both holomorphic or both antiholomorphic indices,
$I_M^{\ N}   =  \{i\delta_m^{n},   -i\delta_{\bar m}^{\bar n} \}$.\footnote{The inverse problem: to f\/ind complex coordinates once the complex structure matrix is given
is nontrivial and does not always have a solution. The manifolds that are not complex but where the matrix
$I_M^{\ N}$ with the properties~(\ref{I}) exists are called {\it almost complex} manifolds. For such manifolds, one cannot
def\/ine a nilpotent holomorphic exterior derivative operator (there is no supersymmetry)
and hence one cannot def\/ine Dolbeault complex and its
index.}

 Consider the standard Christof\/fel symbols
 \begin{gather}
\label{kris}
\Gamma^P_{MN}   =  \frac 12 g^{PQ} \left(   \partial_M g_{PN} + \partial_N g_{PM}  -\partial_Q g_{MN} \right)
   \end{gather}
and the corresponding covariant derivative operator $\nabla_M$. Nonvanishing components of $\Gamma^P_{MN}$ for the Hermitian metric
(\ref{metric}) are
  \begin{gather*}
\Gamma^p_{mn}   = \left( \Gamma^{\bar p}_{\bar m \bar n} \right)^\star =
  \frac {h^{\bar q p} }2 \left(\partial_m h_{n \bar q} + \partial_n h_{m \bar q} \right), \nonumber\\
\Gamma^{\bar p}_{n \bar m} = \Gamma^{\bar p}_{\bar m  n}   = (\Gamma^{p}_{m \bar n} )^\star  = (\Gamma^{p}_{\bar n m} )^\star   =
\frac
{h^{\bar p q}} 2 (\partial_n h_{q \bar m} - \partial_q h_{n \bar m} )  .%\label{krisHerm}
  \end{gather*}
If the metric is K\"ahler, the components of mixed holomorphicity vanish and only the components~$\Gamma^p_{mn}$ and~$ \Gamma^{\bar p}_{\bar m \bar n} $ survive. This implies that the complex structure tensor is covariantly constant, $\nabla_P I_M^{\ N} = 0$.
(The condition $\nabla_P   g_{MN} = 0$ is also, of course, satisf\/ied.)

Introduce the vielbeins $e^A_M$,  $e^M_A$  (such that $e^A_M e^M_B = \delta^A_B$ and $e^A_M e^A_N = g_{MN}$)
and consider the spin connection
  \begin{gather}
\label{Omega}
 \Omega_{M, A}^{\ \ \ \  B} = e_{AN} \big( \partial_M e^{BN} + \Gamma^N_{MK} e^{BK} \big)
\end{gather}
and the corresponding 1-form $\Omega_A^{\ B} =  \Omega_{M, A}^{\ \ \ \ B} dx^M$. It satisf\/ies the standard
Maurer--Cartan structure equation
 \begin{gather*}
%\label{Cartan1}
de_A + \Omega_{A}^{\ B} \wedge e_B = 0 .
\end{gather*}
For K\"ahler metrics, the only nonvanishing components of $\Omega_A^{\ B}$ are $\Omega_a^{\ b}$ and $\Omega_{\bar a}^{\ \bar b}$. The same
concerns the curvature matrix (whose elements represent 2-forms)
 \begin{gather*}
%\label{RAB}
R_A^{\ B}   =  d \Omega_A^{\ B} + \Omega_A^{\ C} \wedge \Omega_C^{\ B},
 \end{gather*}
where the mixed components $R_a^{\ \bar b}$ and $R_{\bar a}^{\ b}$ vanish.
 This means that the holomorphic tangent vectors stay holomorphic after a parallel transport along a closed contour: the holonomy group
is $U(n)$.

If the manifold is not K\"ahler, $\nabla I \neq 0$, the mixed components in $\Omega_A^{\ B}$ and  $R_A^{\ B}$ survive, and the holonomy group is
$SO(2n)$ as for a generic $2n$-dimensional manifold. An important remark is, however, that one {\it can} make $I$ covariantly constant
and reduce the holonomy group to $U(n)$ {if allowing for nonzero torsions}.

 Consider an {\it affine connection}
 \begin{gather*}
%\label{gamK}
\gamma^P_{MN}   =  \Gamma^P_{MN} + K^P_{MN}    ,
 \end{gather*}
where $\Gamma^P_{MN}$ is written in (\ref{kris}) and $K^P_{MN}$ is a {\it tensor} (its tensor nature
will be important for us in the following) called {\it contorsion} tensor.
For the metric to be still covariantly constant, $K_{Q, MN} = g_{PQ} K^P_{MN}$ should satisfy the condition
$K_{Q, MN} = - K_{N, MQ}$. This implies a nonzero {\it torsion}
 \begin{gather*}
%\label{torsion}
T^P_{MN} = \gamma^P_{MN} - \gamma^P_{NM} =  K^P_{MN} - K^P_{NM} \neq 0   .
 \end{gather*}
Let us impose an additional requirement for the tensor $C_{QMN} = g_{PQ} T^P_{MN}$ to be totally antisymmetric.
Then, for a complex manifold, there is a  {\it unique} af\/f\/ine connection
 \begin{gather}
\label{gamC}
  \hat {\Gamma}^P_{MN}   =  \Gamma^P_{MN} + \frac 12 g^{PQ} C_{QMN}
 \end{gather}
satisfying $\hat{\nabla} g = \hat{\nabla} I = 0$.\footnote{There are {\it many} such
connections if the condition $C_{QMN} = C_{[QMN]}$ is not imposed. One of them is the so called
{\it Chern} or {\it Hermitian connection} with the only nonzero components
$\Gamma^C_{\bar p, nm} = \left( \Gamma^C_{p, \bar n \bar  m} \right)^* = \partial_n h_{m\bar p}$.}
 The components of $C_{QMN}$ are expressed through the metric as
\begin{gather}
C_{j k\bar l} = - C_{j\bar l k}  = C_{\bar l jk} = \partial_{k} h_{j\bar l} - \partial_{j} h_{k\bar l}   ,\nonumber\\
C_{\bar j \bar k  l} = - C_{\bar j  l \bar k  } = C_{l \bar j \bar k} =
 (C_{j k \bar l})^* = \partial_{\bar k} h_{l \bar j} - \partial_{\bar j} h_{l \bar k} .\label{Cikl}
\end{gather}
The expressions (\ref{Cikl}) keep their form under holomorphic coordinate transformations. One can also represent
the torsions in an arbitrary frame in real notations
 \begin{gather}
\label{Creal}
C_{QMN} = I_Q^{\ P} I_M^{\ R} I_N^{\ T}  \left( \nabla_P I_{RT} +
\nabla_R I_{TP} + \nabla_T I_{PR} \right)
  \end{gather}
such that the tensor nature of $C_{QMN}$ is seen explicitly.

 The connection (\ref{gamC}), (\ref{Creal})
arises naturally when a supersymmetric
 Lagrangian describing the Dolbeault complex on a generic complex manifold
is built up \cite{my}. Mathematicians know it as the {\it Bismut connection}.
The corresponding Bismut spin connection (it is given by (\ref{Omega}) with~$\hat{\Gamma}$ substituted for~$\Gamma$) and the Bismut curvature matrix {\it do} not have mixed components such that the holonomy group with this curvature
is $U(n)$, as for the K\"ahler manifolds.

When the torsion is present, the Riemann tensor $R_{MNPQ}$ is not symmetric anymore with respect to interchange $\{MN\} \leftrightarrow
\{PQ\} $. We have instead
  \begin{gather*}
  R_{MNPQ}(T)   =  R_{PQMN}(-T)   .
 \end{gather*}

\section{The SQM model}

Following \cite{my}, consider the
chiral superf\/ields
   \begin{gather*}
Z^j = z^j +\sqrt{2}\theta \psi^j - i\theta\bar\theta \dot{z}^j,
\qquad \bar Z^{\bar{j}} = \bar z^{\bar{j}}
-\sqrt{2}\bar\theta \bar\psi^{\bar{j}} +
i\theta\bar\theta \dot{\bar{z}}^{\bar{j}},
   \end{gather*}
$j, \bar j = 1,\ldots,n$,  $\bar D Z = D \bar Z = 0$,
 and choose the supersymmetric action in the following form
\begin{gather}
\label{start}
 S = \int dt d^2\theta \left( -\frac{1}{4}
h_{j\bar k}(Z, \bar Z)  D Z^j \bar D\bar Z^{\bar k}  +
   W(Z, \bar Z) \right)
\end{gather}
with Hermitian $h_{j \bar k}$.

The component Lagrangian of this model can be cast in the following nice form
\begin{gather}
L   =  \frac{1}{2}\left[ g_{MN} \dot x^M \dot x^N + ig_{MN}\psi^M \hat{\nabla} \psi^N
- \frac{1}6 \partial_P C_{MNT}\psi^P\psi^M\psi^N\psi^T \right] \nonumber\\
\phantom{L=}{} + A_M \dot{x}^M  - \frac i2 F_{MN} \psi^M \psi^N ,  \label{1comp}
\end{gather}
where $x^M \equiv \{z^j, \bar z^{\bar j} \}$, etc.\ and the torsion tensor $C_{MNT}$ is given in
(\ref{Cikl}), (\ref{Creal}).
 The second line in (\ref{1comp}) involves  the gauge potential
$A_M = -I_M^{\ N} \partial_N W = ( -i\partial_m W, i\partial_{\bar m} W)$ and its f\/ield strength
$F_{MN} = \partial_M A_N - \partial_N A_M$.

The Lagrangian (\ref{1comp}) is the Lagrangian of some particular supersymmetric sigma-model, its bosonic part describing
 free motion over the manifold.
There are two complications here compared to the K\"ahler case: $(i)$~the covariant derivative
$\hat{\nabla} \psi^M = \dot{\psi}^M + \hat \Gamma^M_{NK} \dot{x}^N \psi^K$   involves now the Bismut connection rather than usual torsion-free connection; $(ii)$~a~4-fermion term is present.

We will discuss the 4-fermion term a bit later, but let us f\/irst notice the appearance of the modif\/ied torsionfull
connection. What is rather nontrivial and somewhat confusing is the fact that the  supercharges also
involve some modif\/ied connections, but this modif\/ication is {\it not the same} as in the Lagrangian!

The quantum covariant supercharges
were presented in \cite{my} in the form
 \begin{gather}
  Q = \sqrt{2}\psi^c e^k_c\left[\Pi_k -\frac{i}{2} \partial_k  (\ln \det \bar e)
- i \psi^b \bar\psi^{\bar a}\Omega_{k, b}^{\ \ \ a} \right],\nonumber\\
  \bar Q = \sqrt{2}\bar\psi^{\bar c} e^{\bar k}_{\bar c}
\left[\bar\Pi_{\bar k} -\frac{i}{2} \partial_{\bar k} (\ln \det e)
+ i \bar\psi^{\bar b} \psi^{a} \bar{\Omega}_{\bar k, a}^{\ \ \ b}\right]   , \label{Qcovgen}
\end{gather}
where $\det  e$ and $\det  \bar e$ are the determinants of the holomorphic and antiholomorphic vielbein matrices
$e_k^a$, $e_{\bar k}^{\bar a}$,
$\Pi_k = -i(\partial_k - \partial_k W)$, and $\bar \Pi_k = -i(\partial_{\bar k} + \partial_{\bar k} W)$.
This expression involves the ordinary torsionless connections $\Omega$, but not all components of the latter,
only the holomorphic ones\footnote{The quantum supercharges and the quantum Hamiltonian  (the expression for the latter was derived in \cite{my},
 but we do not need it here) act on the holomorphic wave functions
\begin{gather*}
%\label{wave}
\Psi(z^k, \bar z^k; \psi^a)   =  C^{(0)}(z^k, \bar z^k) + \psi^j C^{(1)}_j (z^k, \bar z^k) + \cdots
+ \psi^{j_1}\cdots \psi^{j_n}C^{(n)}_{j_1 \dots j_n}(z^k, \bar z^k)
 \end{gather*}
($\psi^j = e^j_a \psi^a$). The coef\/f\/icients in this expansion are isomorphic to the forms~(\ref{Ap}).}.

When
\begin{gather}
\label{W}
 W = W_0 = \frac 12 \ln \det \bar e   ,
 \end{gather}
the operator $Q$ is isomorphic to the external holomorphic derivative $\partial$, and $\bar Q$~-- to~$\partial^\dagger$.

For other choices of $W$, we are dealing with the {\it twisted} Dolbeault complex involving an extra gauge
f\/ield, $\partial X \to \partial X - iA' \wedge X$ with
$A' =   -i \partial_j (W-W_0) dz^j + i \partial_{\bar j} (W-W_0) d\bar z^{\bar j} $.

Remarkably, one can also derive
\cite{Braden,Mavra,FIS} that the { sum} $Q + \bar Q$ is isomorphic
to the Dirac operator $ i  \nabla \!\!\!\!/ =
i{\tilde \nabla}_M  \gamma^M$  with ${\tilde \nabla}_M = \partial_M - iA_M  +
(1/4) {\tilde \Omega}_{M, AB} \gamma^A \gamma^B$
involving the gauge f\/ield $A = A' + A^{(0)}$, where{\samepage
 \begin{gather}
\label{detbundle}
A_M^{(0)}  = \frac i2 \{ -\partial_m \ln \det \bar e,
\partial_{\bar m} \ln \det e \}   .
  \end{gather}
The potential (\ref{detbundle}) is gauge equivalent to  $\frac i4 \{-\partial_m , \partial_{\bar m} \} \ln \det h$.\footnote{Mathematicians call the shift $A^{(0)}$ which appears when establishing this Dolbeault $\leftrightarrow$ Dirac
correspondence
 the connection of the {\it determinant bundle}
(to be quite precise, its {\it square root}).}}

Besides the gauge f\/ield, ${\tilde \nabla}_M$ involves also the spin  connection
\begin{gather}
\label{tildeOmega}
{\tilde \Omega}_{M, AB} = \Omega_{M, AB} - \frac 16 e_A^L e_B^K C_{MLK}   ,
 \end{gather}
    where the torsions appear with the extra factor $1/3$ compared to the Bismut connection entering the Lagrangian.
The dif\/ference $S = Q - \bar Q$ is isomorphic to the operator $\gamma^M I_M^{\ N} {\tilde \nabla}_N$.

The presence of the extra supercharge $S$ and hence the presence of the { new} supersymmetric structure $\{ i \nabla \!\!\!\!/ ,
S \}$  in addition to the well-known chiral structure
$ \{i \nabla \!\!\!\!/ ,  \nabla \!\!\!\!/ \, \gamma^{D+1} \}$ ($\gamma^{D+1}$ being the multidimensional
analog of $\gamma^5$) was noticed f\/irst in \cite{Wipf} for K\"ahler manifolds. But as we see now, this new structure
is present for all complex manifolds with the only complication that  the Dirac operator
and  the operator $S$ involve now the torsionfull connections~(\ref{tildeOmega}).

\section{The index}

From now on, the manifold is assumed to be compact (for noncompact manifolds, the spectrum
of the Hamiltonian is continuous and the notion
of the index is ill-def\/ined)\footnote{See, however, \cite{S4} where the Dolbeault complexes on
$S^4\backslash \{\cdot\}$ and $S^6\backslash \{\cdot\}$ were studied.
In spite of that $S^4\backslash \{\cdot\}$   is not compact, the spectrum of the Dolbeault Laplacian is still discrete, if including
in the Hilbert space square integrable functions with the factor $\sqrt{g} = 1/(1 + \bar z^j z^j)^4$ in the measure.}.
 As we have just seen, the problem of calculating the index of the Dolbeault complex\footnote{Mathematicians sometimes call this index {\it arithmetic genus}.}
is reduced to the problem of calculating the index of the ``nonstandard''
Dirac operator involving extra torsions. The index of the ordinary Dirac operator was, of course,
 calculated by Atiyah and Singer and then in~\cite{A-GFW} using physical functional integral methods. In application to complex manifolds,
this calculation was recently discussed in some details (the derivation in original papers
was rather sketchy)  in~\cite{my}. The Witten index of our SQM system is expressed via the path integral
 \begin{gather*}
I =  {\rm Tr} \big\{ (-1)^F e^{-\beta H} \big\}  =    \lim_{N \to \infty}   \int \prod_\tau  \det h(\bar z^{\bar j}(\tau),
z^j(\tau)) \prod_{j} \frac {d\bar z^{\bar j}(\tau) dz^j(\tau)}{2\pi (\beta/N) }
\nonumber\\
\hphantom{I =  {\rm Tr} \big\{ (-1)^F e^{-\beta H} \big\}  = }{}
\times  \prod_{a} d\psi^{ a}(\tau) d\bar\psi^{\bar a}(\tau)
 \exp \left\{ - \int_0^\beta L_E(\tau) d\tau \right \},%\label{poslepi}
 \end{gather*}
where $L_E(\tau)$ is the Euclidean Lagrangian, $N$ is the number of points into which the Eucli\-dean time interval
$(0,\beta)$ is subdivided, and the periodic boundary conditions, $z(\beta) = z(0)$, \mbox{$\psi(\beta) = \psi(0)$}, are
imposed onto all f\/ields. The integral does not depend on $\beta$. To calculate it, we consider the semiclassical limit $\beta \ll 1$
when the integral is saturated by constant or nearly constant f\/ields. For most systems, one can assume the
f\/ields to be constant, neglect higher
Fourier harmonics and trade
the functional integral for the ordinary one~\cite{Cecotti}. However, in this problem, such a simplif\/ied procedure does not
work. One has to take
into account higher harmonics which amounts to calculating loops.

For K\"ahler metric, it is suf\/f\/icient to perform a one-loop calculation. An accurate analysis (see~\cite{my} for details) shows that
one loop contributions are of the same order as the tree level ones. Even though
 the former include a formally small factor $\beta$, but this factor
is always
multiplied by the structure $\bar \psi^{(0)} \psi^{(0)}$  ($\psi^{(0)}$ being the zero Fourier harmonic of the fermion f\/ield), which is of
order of $1/\beta$, as is seen from the tree level integral over zero harmonics. Now, {\it for K\"ahler manifolds}, one can show that the
second and higher loop contributions are suppressed. The explicit one-loop calculation gives the result
   \begin{gather}
\label{indAS}
I   =  \int e^{{\cal F}/2\pi} {\det}^{-1/2} \left[ \frac {\sin \frac {\cal R} {4\pi}}  { \frac {\cal R} {4\pi}} \right],
 \end{gather}
 where ${\cal F}$ is the gauge f\/ield strength 2-form and ${\cal R}$ is the matrix
2-form associated with the Riemann tensor
   \begin{gather}
\label{FR}
 {\cal F} = \frac 12 F_{MN}\, dx^M \wedge dx^N   , \qquad {\cal R}_A^{\ B} = \frac 12 R_{A\  MN}^{\ B}\, dx^M \wedge dx^N   .
 \end{gather}
 When expanding the integrand in (\ref{indAS}) in Taylor series around unity,
it represents a superposition of forms of dif\/ferent dimensions, but one has, of course, to pick up
only the terms involving the top form $\propto dx^1 \wedge \cdots \wedge dx^{2n}$. Otherwise, the integral is zero.
The determinant factor appearing in the integrand just indicates that we have
performed a one-loop calculation with integrating over the higher Fourier modes.

In the generic complex case, the situation is substantially complicated by the presence of the 4-fermion term in (\ref{1comp}).
As $\bar \psi \psi \sim 1/\beta$, the integral of the 4-fermion term is estimated to be $\int_0^\beta (\psi^4)d\tau
 \sim \beta/\beta^2 \sim \beta^{-1}$.
For small~$\beta$, this contribution is {\it large}! Of course, being a total derivative, it does not
contribute to the integral at the tree level
but, after doing loop integrals, it could in principle be multiplied by some other structure and give a nonvanishing contribution. Actually,
counting
the powers of $\beta$ displays a worrisome fact that, to obtain a reliable result, we should perform in this
case a honest two-loop calculation in~4 and~6 dimension, a honest three-loop calculation in~8 and~10 dimensions, etc.

Obviously, this calculation is not easy. A two-loop calculation for the heat kernel
(the analog of the K\"ahler heat kernel
$e^{{\cal F}/2\pi} \det^{-1/2}[\cdots]$ in equation~(\ref{indAS})) of a 4-dimensional torsionfull Dirac operator was performed
in~\cite{Waldron}
(and, indeed, this heat kernel involves rather intricate total derivative contributions). But nothing is known for higher
dimensions.

Note, however, that there is a class of non-K\"ahler manifolds, namely, the manifolds where the form
$\omega = h_{j \bar k} \, dz^j \wedge d\bar z^{\bar k} $  though not closed,  satisf\/ies the condition $\partial \bar \partial \omega  = 0 $
such that the 4-fermion term vanishes. These manifolds are called SKT manifolds\footnote{SKT stands for strong K\"ahler with torsion. A little bit confusing because these manifolds {\it are} not K\"ahler, but, anyway,
 this is how they are called.}. It is known,
for example, that all non-K\"ahler complex manifolds of complex dimension 2 ({\it complex surfaces})
belong to this class (more exactly, for any complex surface a SKT metric can always be chosen~\cite{Gaud}).
In this case, the functional integral for the Dolbeault index can be calculated by the same token  as for the torsionless Dirac operator.
We obtain
   \begin{gather}
\label{indBis}
I   =  \int e^{({\cal F}' +  {\cal F}_0)/2\pi}   {\det}^{-1/2} \left[ \frac {\sin \frac {\hat{\cal R}} {4\pi}}
 { \frac {\hat{\cal R}} {4\pi}} \right],
 \end{gather}
where ${\cal F}_0$ corresponds to the ``geometrically induced''
potential (\ref{detbundle}), ${\cal F}' = dA'$, with $A'$ being the extra gauge f\/ield of the twisted Dolbeault complex,
and $\hat{\cal R}$ is the curvature form of the Bismut connection that enters the Lagrangian~(\ref{1comp}).
This formula was obtained in~\cite{Bismut} by a~rather ref\/ined mathematical
reasoning. We see that the functional integral method allows one to derive it immediately for almost no price.

However, the formula~(\ref{indBis}) for the Dolbeault index is not the way it was represented in~\cite{Hirz,AS}.
They wrote instead
   \begin{gather}
\label{indpoAS}
I   =  \int e^{{\cal F}'/2\pi} \, {\rm Td}(TM) ,
 \end{gather}
where the symbol Td$(TM)$ ({\it Todd class of a complex tangent bundle} associated with the mani\-fold~$M$) is spelled out as
 \begin{gather}
\label{Todd}
 {\rm Td}(TM)   =
  \prod_{\alpha = 1}^n \frac {\lambda_\alpha/2\pi}{1 - e^{-\lambda_\alpha/2\pi} }  ,
  \end{gather}
where $\lambda_\alpha$ are eigenvalues of the curvature matrix.

This {\it is} the Hirzebruch--Riemann--Roch theorem\footnote{It represents the multidimensional generalization of the {\it Riemann--Roch theorem}. The latter is a statement about the
dimensions
of the moduli spaces of  meromorphic functions on Riemann surfaces, but this statement can be shown to be
equivalent to the statement (\ref{indpoAS})
 about the Dolbeault index of Riemann surfaces equipped by Abelian gauge f\/ields. (If trying to squeeze such a f\/ield on a single map,
singularities (Dirac strings) appear. This singularities may be associated with the poles of meromorphic functions.)
The HRR theorem was originally proven by Hirzebruch only for K\"ahler manifolds, but Atiyah and Singer proved it
for all complex manifolds.}.

One must say that there is a signif\/icant confusion associated with this formula in the physical literature.
 The  curvature matrix is def\/ined in (\ref{FR}). Were the elements of ${\cal R}_A^{\ B}$ ordinary numbers, the eigenvalues
can be found by diagonalizing
 \begin{gather}
\label{diag}
{\cal R} \to {\rm diag} (i\lambda_1 \sigma_2, \ldots, i \lambda_n \sigma_2 )
 \end{gather}
($\sigma_2$ being a Pauli matrix). They are not numbers, however, but 2-forms. Therefore,
the eigenvalues $\lambda_\alpha$ do not have as such a lot of
meaning and only their certain combinations (like $\sum\limits_{\alpha =1}^n \lambda_\alpha^2 = {\rm Tr} \{ {\cal R}\wedge
{\cal R} \}$)
may have.

For K\"ahler manifolds, mixed holomorphic components ${\cal R}_a^{\ \bar b}$, ${\cal R}_{\bar a}^{\ b}$ vanish and
one can consider instead of the antisymmetric $2n \times 2n$ matrix
${\cal R}_{AB}$, the $n \times n$ matrix
${\cal R}_a^{ \ b}$ and {\it its} eigenvalues\footnote{Note that, with the convention (\ref{diag}) chosen,
the eigenvalues of ${\cal R}_a^{\ b}$ are not $\lambda_\alpha$, but $-i\lambda_\alpha$.}.
The symmetric polynomials
of the roots of the characteristic equation (having the order $n$), which enter equation~(\ref{Todd}),
are expressed in a simple way into its coef\/f\/icients and are reduced to combinations  of products of the invariants
${\rm Tr}\{{\cal R}\}$, ${\rm Tr}\{{\cal R} \wedge R\}$, etc.
But in a generic non-K\"ahler case, the characteristic equation has the order  $2n$. It involves only even powers of $\lambda$,
such that its roots come in pairs $\{\lambda_\alpha, -\lambda_\alpha \}$, and an expression
like $\sum\limits_{\alpha=1}^n \lambda_\alpha$ is simply meaningless!

This is true for the ordinary Riemann tensor that does not respect the holomorphic structure in the tangent space.
We have learned, however, that even if the manifold is not K\"ahler, one can def\/ine the Bismut connection (\ref{gamC}), (\ref{Cikl})
and the corresponding curvature tensor which {\it do} respect the holomorphic structure.  The eigenvalues $\lambda_\alpha$ entering
equation~(\ref{Todd}) refer to {\it this} Bismut torsionfull curvature tensor, not the usual one\footnote{To be more precise, one needs not to consider necessarily a Bismut connection, any connection satisfying $\nabla g = \nabla I = 0$ is {\it a}
complex tangent bundle connection and can be used to calculate the index. (To reiterate, the ordinary {\it real} tangent bundle
connection~(\ref{Omega}) {\it can}not be used for this purpose.)
Even though the Todd polynomials~(\ref{Todd}) of dif\/ferent such complex connections are dif\/ferent, their integrals
(\ref{indpoAS}) are the same. We will see it soon.}!

One can show that, in the K\"ahler case, the integrands in (\ref{indBis}) and (\ref{indpoAS}) coincide.
Indeed, the factor $\det{}^{-1/2}[\cdots]$ in (\ref{indBis}) can be represented as\footnote{The r.h.s.\ of equation~(\ref{Aroof}) is called {\it A-roof genus}.}
 \begin{gather}
\label{Aroof}
{\det}^{-1/2} \left[ \frac {\sin \frac {\cal R} {4\pi}}  { \frac {\cal R} {4\pi}} \right]
=\ \prod_{\alpha = 1}^n \frac {\lambda_\alpha/4\pi}{\sinh   (\lambda_\alpha/4\pi) }  ,
 \end{gather}
while
\begin{gather}
\label{F0sumlam}
e^{{\cal F}_0/2\pi} \   \stackrel{\text{K\"ahler}}{=} \  e^{i{\cal R}_a^{\ a} /4\pi}
  =  \exp\left\{ \frac 1{4\pi} \sum_\alpha \lambda_\alpha \right\}   .
 \end{gather}
The identity (\ref{Aroof}) is valid in a generic complex case (one has only to substitute $\hat {\cal R}$ for ${\cal R}$),
but the relation~(\ref{F0sumlam}) is not.

In the generic case
 \begin{gather}
\label{dobavka}
i\hat {\cal R}_a^{\ a}   =  id\hat \Omega_a^{\ a} = i d \left[ \left(\partial_{\bar j} \ln \det e + \frac 12 h^{\bar k t } C_{\bar j \bar k t}\right) d\bar z^{\bar j}
- \left(\partial_j \ln \det \bar e + \frac 12 h^{\bar t k} C_{jk\bar t} \right)dz^j
 \right]  ,
\end{gather}
i.e.\ the connection (\ref{detbundle}) (that would lead alone to $\sum_\alpha \lambda_\alpha$ )is shifted by a certain vector.

Following the logics of \cite{WuZee}, it is not dif\/f\/icult to see, however, that this shift does not af\/fect the value
of the integral. The variation of the integrand in~(\ref{indBis}) under an inf\/initesimal shift $A^{(0)} \to A^{(0)} + \epsilon B$
involves various terms.
Consider one of them, say, the term
\[
 X   =   \epsilon   dB\wedge {\cal F}_0 \wedge \hat {\cal R}_a^{\ b} \wedge \hat{\cal R}_b^{\ a}   .
 \]
Due to the facts $d{\cal F}_0 = 0$ and $ d (\hat {\cal R}_a^{\ b} \wedge \hat {\cal R}_b^{\ a}) = 0$ (the latter identity is proven using the Bianchi
identities
\[
d\hat {\cal R}_a^{\ b} - \hat {\cal R}_a^{\ C}\wedge \hat \Omega_C^{\ b} + \hat\Omega_a^{\ C}\wedge \hat
{\cal R}_C^{\ b}   =  0
\]
 and the fact that, for
the Bismut connection, only the holomorphic dummy indices $C=c$ contribute), $X$ is an exact form and its integral is zero.
 The same reasoning can be applied to all other terms in the variation, like
\[
Y = \epsilon   dB\wedge {\cal F}_0 \wedge {\cal F}' \wedge {\cal F}' \wedge {\rm Tr} \, {\hat {\cal R}^4} \wedge {\rm Tr} \, {\hat {\cal R}^6}
= \epsilon   d\big(B\wedge {\cal F}_0 \wedge {\cal F}' \wedge {\cal F}' \wedge {\rm Tr}\, {\hat {\cal R}^4} \wedge {\rm Tr}\, {\hat {\cal R}^6}\big)   ,
\]
etc.

Consider now a f\/inite shift $A(t) = A + tB$.  We have just seen that the derivative of the integral over $t$ is zero. This means that
$I(t)$ does not depend on $t$, i.e.\ it is the same for the shifted and unshifted gauge f\/ield.

This proof relies on the fact that $B$ is a regular 1-form well def\/ined at every point of our manifold and satisfying the condition
(\ref{norm}). Were it  not so, for example, if $B \propto A$, the variation would still  formally be an exact form, but it would
involve singularities, and its integral might not vanish.

This proves the HRR theorem for the SKT manifolds. How to proceed in a general case when the 4-fermion term in (\ref{1comp})
does not vanish and we cannot calculate the path integral?

\looseness=1
We have seen that the problem of evaluating the Dolbeault index is reduced to the problem of evaluating the index of
 a nonstandard torsionfull
Dirac operator. The point is that one can {\it deform} this nonstandard operator such that the torsions disappear. The index, however,
is not changed under such smooth deformation. The invariance of a topological index under smooth deformation is a tool widely used
in physics to calculate Witten indices of dif\/ferent supersymmetric theories \cite{Witind}. It always works under the condition
that supersymmetry survives under such deformation\footnote{And if supersymmetry does not survive, it does not work. Torsions representing regular tensors on a manifold
can be unwinded. But  topologically nontrivial
gauge f\/ields cannot. Indeed, even if disregarding the fact that a~correctly from mathematical viewpoint
 def\/ined f\/iber bundle has always an integer
topological charge and just trying to solve the Schr\"odinger equation for a~system with a fractional
topological charge (for example, a fractional magnetic f\/lux on~$S^2$), one would see that, for fractional charges, supersymmetry is lost~\cite{flux}. And this is the reason why the deformation philosophy does not work in this case.}.

\looseness=1
Does the supersymmetry survive under an arbitrary torsion shift? After all, we have seen that the
Dolbeault operator is isomorphic
to the Dirac operator with some {\it particular}
 torsions given in~(\ref{Cikl}). For other torsions, Dolbeault supersymmetry
encoded in the superspace action~(\ref{start}) and associated with the existence of two
supercharges  $i{\tilde \nabla}_M  \gamma^M$
and $S = \gamma^M I_M^{\ N} {\tilde \nabla}_N$ is lost. The point is, however, that an even-dimensional
Dirac operator, irrespectively of whether it involves torsions or not,  enjoys
the supersymmetry associated with the chiral pair
of supercharges  $ \{i \nabla \!\!\!\!/ $~and~$  \nabla \!\!\!\!/ \, \gamma^{D+1} \}$. The latter
 is true under two
conditions:  $i)$ The manifold should admit {\it spin structure}\footnote{To be quite precise, spin$^C$ structure.}, which means that
topological charge associated with the gauge f\/ield
is quantized in a proper way (cf.\ the footnote remark above). $ii)$ The torsions should represent a regular tensor.
If either of these conditions is violated,
the result of the action of $ i \nabla \!\!\!\!/ $ on the regular spinors might be
singular, and this breaks supersymmetry.

But in our case, the torsions (\ref{Creal}) are regular tensors and the
gauge f\/ield topological charge is quantized as it should.
We then conclude that {\it the Dolbeault index for an arbitrary
compact complex manifold coincides with the index of the associated Dirac operator with unwinded torsions}.
This gives the following result
 \begin{gather}
\label{indSmil}
I   =  \int e^{{\cal F}'/2\pi} \exp \left\{ \frac 1{16\pi}
I_M^{\ P}   \partial_N \partial_P (\ln \det g)   dx^M \wedge dx^N \right\}
{\det}^{-1/2} \left[ \frac {\sin \frac {\cal R} {4\pi}}  { \frac {\cal R} {4\pi}} \right],
 \end{gather}
where we have explicitly written in real notations the contribution due to the metric-indu\-ced gauge f\/ield
(\ref{detbundle}).
The determinant factor involves the conventional torsionless Riemann tensor.

Let us show now that the integral (\ref{indSmil}) coincides with (\ref{indpoAS}).
Again, we can use the reasoning of \cite{WuZee}.

Consider the variation of the integrand in (\ref{indSmil}) under an inf\/initesimal deformation
\[
\omega_A^{\ B}    \to  \omega_A^{\ B} + \epsilon \lambda_A^{\ B}   ,
\]
where $\lambda_{PMN} = e^A_M e_{BN}\lambda_{P, A}^{\ \ \ B}$ is a {\it tensor}. Then, say, the structure ${\cal R}_A^{\ B} \wedge {\cal R}_B^{\ A}$ is shifted as
\[
\delta\big({\cal R}_A^{\ B} \wedge {\cal R}_B^{\ A}\big)   =  \epsilon   \big(d\lambda_A^{\ B} + \lambda_A^{\ C} \wedge \Omega_C^{\ B} +
\Omega_A^{\ C} \wedge \lambda_C^{\ B}\big)\wedge {\cal R}_B^{\ A}   =
\epsilon   d\big(\lambda_A^{\ B} \wedge {\cal R}_B^{\ A}\big)   ,
\]
where the Bianchi identity was used. The same reasoning applies to the terms like
${\cal R}_A^{\ B} \wedge {\cal R}_B^{\ C} \wedge {\cal R}_C^{\ D} \wedge {\cal R}_D^{\ A}$, ${\cal F}' \wedge {\cal F}'
\wedge {\cal R}_A^{\ B} \wedge
{\cal R}_B^{\ A} $, etc. In other words, the variation is an exact form and its integral vanishes. This applies also
to a f\/inite deformation: we can deform~(\ref{Omega}) all the way to the Bismut connection and the integral for the index still
does not change.

This brings the integral to the form (\ref{indBis}). Then, following the procedure above, we deform further
the gauge f\/ield~(\ref{detbundle}) as in~(\ref{dobavka})
and bring the integral to the canonical form~(\ref{indpoAS}). The theorem is proven.

\section{Example: Hopf manifolds}

The simplest example for non-K\"ahler complex manifolds are Hopf manifolds.
Their metric has a simple form\footnote{In this section, we do not distinguish between covariant and contravariant, holomorphic and
antiholomorphic indices.}
 \begin{gather}
\label{Hopf}
ds^2   =  \frac {d\bar z_j dz_j}{\bar z_k z_k}   ,
 \end{gather}
and the complex coordinates $z_j$ lie in the region $1 \leq |z| \leq 2$,
with the points $w_j$ and $2 w_j$ being identif\/ied ($|w| = 1$).
The conformally invariant metric (\ref{Hopf}) is consistent
with this identif\/ication. Topologically, the Hopf manifold ${\cal H}^n$ of complex dimension $n$ is a torus
$S^1 \times S^{2n-1}$. The K\"ahler form ${\omega}_n$ associated with the metric (\ref{Hopf}) is not closed and
 ${\cal H}^n$ is not K\"ahler. Note, however, that $\partial \bar \partial   {\omega}_2 = 0$ such that ${\cal H}^2$
represents (as all complex surfaces  do~\cite{Gaud}) an SKT manifold.

The (untwisted) Dolbeault Laplacian acting on the {\it functions} on ${\cal H}^n$ is
 \begin{gather}
\label{LaplDol}
-\frac 12 \triangle^{\rm Dol} \ \stackrel {\text{for  0-forms}}{=}\ - (\bar z_k z_k) \bar \partial_j \partial_j  + z_j \partial_j
 \end{gather}
to be compared with the usual Laplacian
 \begin{gather*}
%\label{lapl}
-\frac 12 \triangle   =  - (\bar z z) \bar \partial  \partial  + \frac 12 (z \partial + \bar z \bar \partial)  .
 \end{gather*}
The operator (\ref{LaplDol}) has one zero mode, $\Psi(\bar z, z) = 1$. Actually, the operator $ \triangle^{\rm Dol}$
gives also zero, when acting on any antiholomorphic function $\chi(\bar z)$. However, if $\chi(\bar z)$ is not constant,
it is not a~function on ${\cal H}^n$ because it is not the same at $\bar z_j = \bar w_j$ and  $\bar z_j = 2\bar w_j$

The full Dolbeault Laplacian has also a zero mode in the sector of (1,0)-forms. To see this, note that
 \begin{gather}
\label{Dolln}
  \triangle^{\rm Dol}   \ln (\bar z z)   =  0  .
  \end{gather}
 $\ln (\bar z z)$ is not a function on ${\cal H}^n$, but the
1-form
 \begin{gather*}
%\label{P}
P   =  \partial \ln (\bar z z)   =  \partial_j [\ln (\bar z z) ] dz_j   = \frac {\bar z_j dz_j }{\bar z_k z_k}
 \end{gather*}
is well def\/ined on ${\cal H}^n$. It satisf\/ies $\partial P = \partial^\dagger P = 0$ and represents thus a zero mode of
the full Dolbeault Laplacian\footnote{The property $\partial P = 0$ is obvious. $\partial^\dagger P = 0$ follows from~(\ref{Dolln}).
One can also check it directly using the map $dz_j \to \psi_j$ and the expressions~(\ref{Qcovgen})
 for the supercharges.}.

The index in this case is equal to $1-1 = 0$. It is instructive to reproduce this result
 by calculating the integrals
(\ref{indpoAS}) or, alternatively, (\ref{indSmil}) (the latter expression is somewhat simpler because
it depends only on the ordinary curvature rather than the Bismut curvature).
 For the Hopf manifold, the symmetry dictates\footnote{The functions $A(x)$ and $B(x)$ can, of course, be calculated, but we do not need them.}
 \begin{gather*}
%\label{RHopf}
 {\cal R}_{MN}   =  A dx_M \wedge dx_N + B (x_M dx_N - x_N dx_M) \wedge (x_Q dx_Q)   .
 \end{gather*}
In addition
\begin{gather*}
%\label{FHopf}
{\cal F} = {\cal F}_0   \propto   \partial_j \bar \partial_{k} \ln(\bar z z) dz_j \wedge d\bar z_{k}   =
\frac {dz_j \wedge d\bar z_j}{\bar z z} - \frac {(\bar z_j dz_j) \wedge (z_k d\bar z_k)} {(\bar z z)^2}  .
 \end{gather*}
One can then be directly convinced that
 \begin{gather}
\label{FFiRR}
{\cal F} \wedge {\cal F} =  0   , \qquad
{\cal R}_{MN} \wedge {\cal R}_{NP}  =  0
 \end{gather}
 for any $A$, $B$.

The index represents a combination of the integrals of the terms like
${\cal F} \wedge {\cal F} \wedge {\cal R}_{MN} \wedge {\cal R}_{NP} \wedge {\cal R}_{PQ} \wedge
{\cal R}_{QM}$ which all vanish due to~(\ref{FFiRR}).

\subsection*{Acknowledgements}
I am indebted to  G.~Carron, E.~Ivanov, L.~Positselsky, S.~Theisen, A.~Wipf and, especially, M.~Verbitsky
for illuminating discussions and remarks.
Special thanks are due to Alexei Rosly for
carefully reading the manuscript and many valuable remarks.

\pdfbookmark[1]{References}{ref}
\LastPageEnding

\end{document}